\documentstyle[11pt,moriond,axodraw]{article}

\newcommand{\lsim}
{\mathrel{\raisebox{-.3em}{$\stackrel{\displaystyle <}{\sim}$}}}

\def\asymp#1%
{\mathrel{\raisebox{-.4em}{$\widetilde{\scriptstyle #1}$}}}

\def\Nequal#1%
{\mathrel{\raisebox{-.5em}{$\stackrel{=}{\scriptstyle\rm#1}$}}}
\newcommand{\dsl}[1]{\not \hspace{-0.7mm}#1}
\def\dsl{\mathpalette\make@slash}
\def\make@slash#1#2{\setbox\z@\hbox{$#1#2$}%
  \hbox to 0pt{\hss$#1/$\hss\kern-\wd0}\box0}

\def\beq{\begin{equation}}
\def\eeq{\end{equation}}
\def\beqar{\begin{eqnarray}}
\def\eeqar{\end{eqnarray}}
\def\barr#1{\begin{array}{#1}}
\def\earr{\end{array}}
\def\bfi{\begin{figure}}
\def\efi{\end{figure}}
\def\btab{\begin{table}}
\def\etab{\end{table}}
\def\bce{\begin{center}}
\def\ece{\end{center}}

\def\text{\textstyle}

\def\al{\alpha}

\def\ga{\gamma}
\def\de{\delta}

\def\si{\sigma}

\def\reffi#1{\mbox{Fig.~\ref{#1}}}

\makeatletter
\def\@citere#1#2{\unskip\ Ref.~#1}
\def\citere#1{{\let\@cite\@citere\cite{#1}}}
\def\@citeres#1#2{\unskip\ Refs.~#1}
\def\citeres#1{{\let\@cite\@citeres\cite{#1}}}
\makeatother

\newcommand{\GeV}{\unskip\,\mathrm{GeV}}
\newcommand{\MeV}{\unskip\,\mathrm{MeV}}

\newcommand{\fb}{\unskip\,\mathrm{fb}}

\newcommand{\ri}{{\mathrm{i}}}
\newcommand{\rd}{{\mathrm{d}}}

\newcommand{\M}{{\cal{M}}}

\def\mathswitchr#1{\relax\ifmmode{\mathrm{#1}}\else$\mathrm{#1}$\fi}

\newcommand{\PW}{\mathswitchr W}

\newcommand{\Pe}{\mathswitchr e}

\newcommand{\Pd}{\mathswitchr d}

\newcommand{\Pu}{\mathswitchr u}
\newcommand{\Pubar}{\bar{\mathswitchr u}}

\newcommand{\Pep}{\mathswitchr {e^+}}
\newcommand{\Pem}{\mathswitchr {e^-}}

\def\mathswitch#1{\relax\ifmmode#1\else$#1$\fi}

\newcommand{\MW}{\mathswitch {M_\PW}}

\newcommand{\GW}{\Gamma_{\PW}}

\newcommand{\GF}{\mathswitch {G_\mu}}

\def\solid{\raise.9mm\hbox{\protect\rule{1.1cm}{.2mm}}}
\def\dash{\raise.9mm\hbox{\protect\rule{2mm}{.2mm}}\hspace*{1mm}}

\newcommand{\eeWW}{{\Pe^+ \Pe^-\to \PW^+ \PW^-}}

\newcommand{\eeWWffff}{\Pep\Pem\to\PW\PW\to 4f}

\newcommand{\eeffff}{\Pep\Pem\to 4f}
\newcommand{\eeffffg}{\eeffff\ga}
\renewcommand{\O}{{\cal O}}

\def\draftdate{\relax}
\def\mda{\relax}
\def\mua{\relax}
\def\mla{\relax}
\def\draft{
\def\thtystars{******************************}
\def\sixtystars{\thtystars\thtystars}
\typeout{}
\typeout{\sixtystars**}
\typeout{* Draft mode!
         For final version remove \protect\draft\space in source file *}
\typeout{\sixtystars**}
\typeout{}
\def\draftdate{\today}
\def\mua{\marginpar[\boldmath\hfil$\uparrow$]%
                   {\boldmath$\uparrow$\hfil}%
                    \typeout{marginpar: $\uparrow$}\ignorespaces}
\def\mda{\marginpar[\boldmath\hfil$\downarrow$]%
                   {\boldmath$\downarrow$\hfil}%
                    \typeout{marginpar: $\downarrow$}\ignorespaces}
\def\mla{\marginpar[\boldmath\hfil$\rightarrow$]%
                   {\boldmath$\leftarrow $\hfil}%
                    \typeout{marginpar: $\leftrightarrow$}\ignorespaces}
\def\Mua{\marginpar[\boldmath\hfil$\Uparrow$]%
                   {\boldmath$\Uparrow$\hfil}%
                    \typeout{marginpar: $\uparrow$}\ignorespaces}
\def\Mda{\marginpar[\boldmath\hfil$\Downarrow$]%
                   {\boldmath$\Downarrow$\hfil}%
                    \typeout{marginpar: $\downarrow$}\ignorespaces}
\def\Mla{\marginpar[\boldmath\hfil$\Rightarrow$]%
                   {\boldmath$\Leftarrow $\hfil}%
                    \typeout{marginpar: $\leftrightarrow$}\ignorespaces}
\overfullrule 5pt
\oddsidemargin -15mm
\marginparwidth 29mm
}

\def\stars{\strut\leaders\hbox{*}\hfill\strut}
\def\starline{\hfil\strut\hfil\hbox to \textwidth {\stars}\hfil}

\begin{document}

%
%
%

\def\thefootnote{\fnsymbol{footnote}}
\setcounter{footnote}{1}
\hfill BI-TP 2000/13 \\
\vspace*{4cm}
\begin{center}
{\Large\bf PRECISE PREDICTIONS FOR W-PAIR PRODUCTION \\[.5em]
AT LEP2 WITH {\sc RacoonWW} 
\footnote{To appear in the 
{\it Proceedings of the XXXVth Rencontres de Moriond, 
Electroweak Interactions and Unified Theories}, March 2000.}
}
\\[1.8cm]
{\large
A.~Denner$^1$,
S.~Dittmaier$^2$%
\footnote{Partially supported by the Bundesministerium f\"ur Bildung und
Forschung, No.~05~7BI92P~9, Bonn, Germany.},
M.~Roth$^3$ and D.~Wackeroth$^4$}\\[1.5em]
\parbox{8cm}{$^1$Paul-Scherrer-Institut, Villigen, Switzerland\\[.5em]
$^2$Universit\"at Bielefeld, Bielefeld, Germany\\[.5em]
$^3$Universit\"at Leipzig, Leipzig, Germany\\[.5em]
$^4$University of Rochester, Rochester NY, USA}
\end{center}
\vspace*{3.0cm}
{\large\bf Abstract}\\[.2cm]
\setlength{\baselineskip}{8pt}
{\sc RacoonWW} is the first Monte Carlo generator for
$\eeWWffff(+\gamma)$ that includes the electroweak ${\cal O}(\alpha)$ 
radiative corrections in the double-pole approximation completely.
Some numerical results for LEP2 energies are discussed, and the
predictions for the total W-pair cross section are confronted with LEP2
data.
\par
\vskip 4cm
\noindent
May 2000        
\null

\clearpage

\def\thefootnote{\fnsymbol{footnote}}
\setcounter{footnote}{1}
\vspace*{4cm}
\begin{center}
{\Large\bf PRECISE PREDICTIONS FOR W-PAIR PRODUCTION \\[.5em]
AT LEP2 WITH {\sc RacoonWW} 
\footnote{Talk presented by S.~Dittmaier.} 
}
\\[.8cm]
A.~Denner$^1$,
S.~Dittmaier$^2$%
\footnote{Partially supported by the Bundesministerium f\"ur Bildung und
Forschung, No.~05~7BI92P~9, Bonn, Germany.},
M.~Roth$^3$ and D.~Wackeroth$^4$\\[.5em]
\parbox{8cm}{$^1$Paul-Scherrer-Institut, Villigen, Switzerland\\
$^2$Universit\"at Bielefeld, Bielefeld, Germany\\
$^3$Universit\"at Leipzig, Leipzig, Germany\\
$^4$University of Rochester, Rochester NY, USA}
\end{center}
\vspace*{1.0cm}
{\large\bf Abstract}\\[.2cm]
\setlength{\baselineskip}{8pt}
{\sc RacoonWW} is the first Monte Carlo generator for
$\eeWWffff(+\gamma)$ that includes the electroweak ${\cal O}(\alpha)$ 
radiative corrections in the double-pole approximation completely.
Some numerical results for LEP2 energies are discussed, and the
predictions for the total W-pair cross section are confronted with LEP2
data.

\normalsize
\vspace{.2cm}

\section{W-pair production at LEP2}

The investigation of W-pair production at LEP2 plays an important role
in the verification of the Electroweak Standard Model (SM). 
Apart from the direct observation of the triple-gauge-boson
couplings in $\eeWW$, the increasing accuracy in the W-pair-production
cross-section and W-mass measurements 
has put this process into the row of SM precision tests. 

To account for the high experimental accuracy \cite{lep2cs,lep2mw}, on the
theoretical side is a great challenge: the W bosons have to be treated
as resonances in the full four-fermion processes $\eeffff$,
and radiative corrections need to be included. While 
several lowest-order
predictions are based on the full set of Feynman diagrams, 
only very few calculations include radiative
corrections beyond the level of universal effects (see 
\citeres{lep2repWcs,lep2mcws} and references therein).
Fortunately, to match the experimental precision for W-pair production at LEP2 
a full one-loop calculation for the four-fermion processes is not needed,
and it is sufficient to take into account only those radiative corrections 
that are enhanced by two resonant W bosons.  
A naive estimate of the neglected corrections yields
$(\al/\pi)(\GW/\MW)\sim 0.5\%$.
The theoretically clean way to carry out this approximation is
the expansion about the two resonance poles, which is called {\it
double-pole approximation} (DPA) \cite{st91}.
A full description of this strategy and of different variants used in
the literature \cite{yfsww,Be98,ku99}
(some of them involving further approximations) is beyond the scope of
this article. We can only briefly sketch the approach pursued in
{\sc RacoonWW} \cite{racoonww_lep2res,De00}.

\section{Radiative corrections with {\sc RacoonWW}}

In DPA, ${\cal O}(\alpha)$ corrections to
$\Pep\Pem\to\PW\PW\to 4f$ can be classified into two types:
factorizable and non-factorizable corrections.
We first focus on virtual corrections.

\begin{figure}
{\centerline{
\unitlength 1pt
\begin{picture}(190,90)(0,10)
\ArrowLine(30,50)( 5, 95)
\ArrowLine( 5, 5)(30, 50)
\Photon(30,50)(150,80){2}{11}
\Photon(30,50)(150,20){2}{11}
\ArrowLine(150,80)(190, 95)
\ArrowLine(190,65)(150,80)
\ArrowLine(190, 5)(150,20)
\ArrowLine(150,20)(190,35)
\GCirc(30,50){10}{.5}
\GCirc(90,65){10}{1}
\GCirc(90,35){10}{1}
\GCirc(150,80){10}{.5}
\GCirc(150,20){10}{.5}
\DashLine( 70,10)( 70,90){2}
\DashLine(110,10)(110,90){2}
\put(40,26){$W$}
\put(40,63){$W$}
\put(120, 6){$W$}
\put(120,83){$W$}
\end{picture}
\hspace*{2em}
\begin{picture}(120,90)(0,10)
\ArrowLine(30,50)( 5, 95)
\ArrowLine( 5, 5)(30, 50)
\Photon(30,50)(90,80){2}{6}
\Photon(30,50)(90,20){2}{6}
\GCirc(30,50){ 7}{0}
\Vertex(90,80){1.2}
\Vertex(90,20){1.2}
\ArrowLine(90,80)(120, 95)
\ArrowLine(120,65)(105,72.5)
\ArrowLine(105,72.5)(90,80)
\Vertex(105,72.5){1.2}
\ArrowLine(120, 5)( 90,20)
\ArrowLine( 90,20)(105,27.5)
\ArrowLine(105,27.5)(120,35)
\Vertex(105,27.5){1.2}
\Photon(105,27.5)(105,72.5){2}{4.5}
\put(93,47){$\gamma$}
\put(55,73){$W$}
\put(55,16){$W$}
\end{picture}
} }
\caption{Structure of virtual factorizable corrections (l.h.s.), with
loop corrections in the blobs, and a
typical diagram for virtual non-factorizable corrections (r.h.s.)}
\label{fig:diags}
\end{figure}
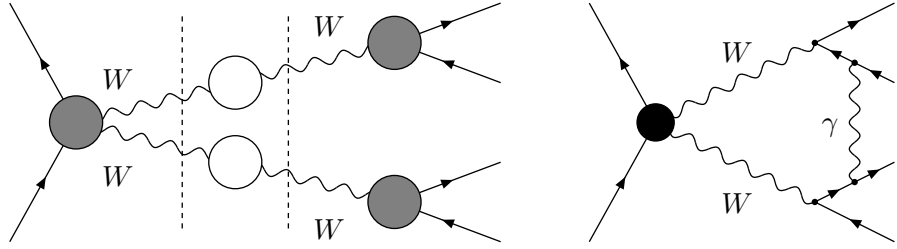

{\it Factorizable corrections} are those that correspond either 
to W-pair production or to W~decay. Virtual factorizable corrections 
are represented by the schematic diagram on the l.h.s.\ of 
\reffi{fig:diags}, in which the shaded blobs contain all one-loop
corrections to the on-shell production and on-shell 
decay processes, and the open blobs
include the corrections to the W~propagators. 
For the corresponding matrix element ${\cal M}$ the application of the
DPA amounts to the replacement
\newcommand{\kp}{k_{\PW^+}}
\newcommand{\km}{k_{\PW^-}}
\beq
{\cal M} = \frac{R(\kp^2,\km^2)}{(\kp^2-\MW^2)(\km^2-\MW^2)} \;\to\;
\frac{R(\MW^2,\MW^2)}{(\kp^2-\MW^2+\ri\MW\GW)(\km^2-\MW^2+\ri\MW\GW)},
\eeq
where the originally gauge-dependent numerator $R(\kp^2,\km^2)$ 
is replaced by the gauge-indepen\-dent residue $R(\MW^2,\MW^2)$.
The one-loop corrections to this residue can be deduced from the known
results for the pair production and the decay of on-shell W~bosons.
However, the spin correlations between the two W~decays should be taken
into account.

{\it Non-factorizable corrections} \cite{be97}
comprise all those doubly-resonant
corrections that are not yet contained in the factorizable ones, 
and include, in particular, all diagrams involving particle exchange between the
subprocesses. Such diagrams only lead to
doubly-resonant contributions if the exchanged particle is a photon
with energy $E_\gamma\lsim\Gamma_\PW$; all other non-factorizable
diagrams are negligible in DPA. A typical diagram for a virtual 
non-factorizable correction is shown on the r.h.s.\ in
\reffi{fig:diags}, where the full blob represents tree-level subgraphs.
We note that diagrams involving photon exchange between the W~bosons
contribute both to factorizable and
non-factorizable corrections; otherwise the splitting into those parts
would not be gauge-invariant.

In {\sc RacoonWW} the virtual corrections are treated in DPA, including
the full set of factorizable and non-factorizable ${\cal O}(\alpha)$ 
corrections. The real corrections are calculated from full 
matrix elements for $\eeffffg$, as described in \citere{ee4fa},
i.e.\ the DPA is not used in this part.
In this way, we avoid potential problems in the definition of the DPA
for the emission of photons with energies $E_\gamma\sim\Gamma_\PW$.
On the other hand, this asymmetry in the calculation of virtual and real
corrections requires particular care concerning the structure of IR and
mass singularities. The singularities have the form of a
universal radiator function convoluted with the
respective lowest-order matrix element squared $|\M_0|^2$ 
of the non-radiative process.  
Since the virtual corrections are calculated in DPA, but the full
matrix element is used for the real photons, a simple summation
of virtual and real corrections would lead to a
mismatch in the singularity structure and eventually to totally wrong
results. Therefore, we extract those singular
parts from the real photon contribution that exactly match the
singular parts of the virtual photon contribution, then replace in
these terms the full $|\M_0|^2$ 
by the one calculated in DPA, and finally add
this modified part to the virtual corrections. This modification is
allowed within DPA and leads to a proper
matching of all IR and mass singularities. This treatment has been
carried out in two different ways, once following phase-space
slicing, once using the subtraction formalism of \citere{subtract}.
\looseness -1

Beyond ${\cal O}(\alpha)$, {\sc RacoonWW} includes soft-photon
exponentiation and leading higher-order ISR effects in the
structure-function approach. Using $\GF$ as input parameter instead of
$\alpha(0)$, also the leading effects from $\Delta\alpha$ and
$\Delta\rho$ are absorbed and partially resummed in the lowest order.
\looseness -1

\section{Phenomenological results}

A survey of numerical results obtained with {\sc RacoonWW} has already
been presented in \citere{racoonww_lep2res} for LEP2 and linear-collider
energies. Here we only review the W~invariant-mass distribution 
given there and extend the results for the total cross section.

\begin{figure}
{\centerline{ 
\setlength{\unitlength}{1cm}
\begin{picture}(7.9,7.3)
\put(0,-.4){\includegraphics{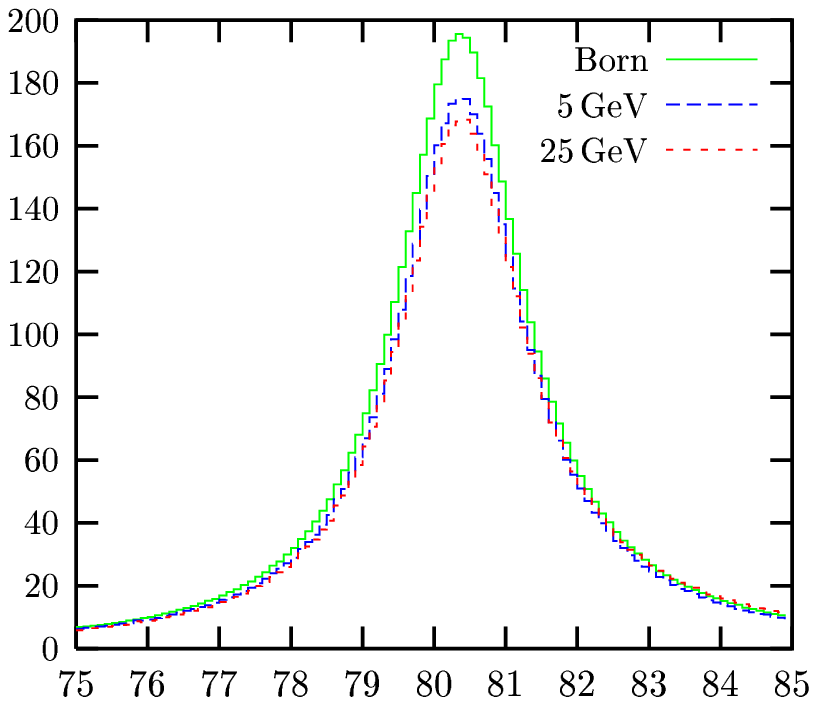}}
\put(0.1,6.6){\makebox(1,1)[l]{$\frac{\rd \si}{\rd M_{\Pd\Pu}}\
\left[\frac{\fb}\GeV\right]$}}
\put(4.0,-0.6){\makebox(1,1)[c]{$M_{\Pd\Pu}\ [\mathrm{GeV}]$}}
\end{picture}%
\begin{picture}(7.9,7.3)
\put(0.5,6.6){\makebox(1,1)[c]{$\de\ [\%]$}}
\put(4.0,-0.6){\makebox(1,1)[c]{$M_{\Pd\Pu}\ [\mathrm{GeV}]$}}
\put(0,-.4){\includegraphics{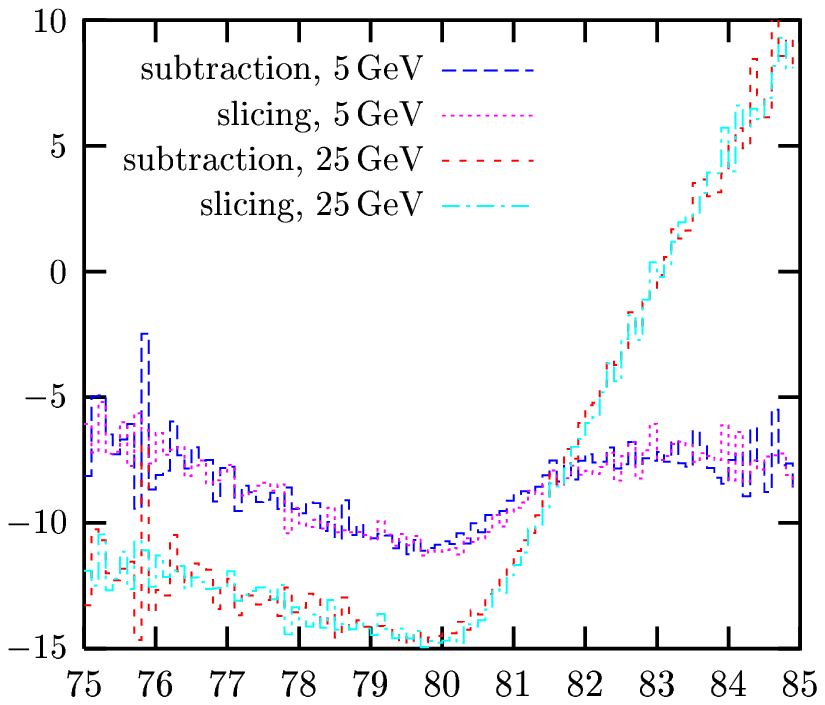}}
\end{picture}
} } 
\caption{Invariant-mass distribution of the $\Pd\Pubar$ pair for
  $\Pep\Pem\to\nu_\mu\mu^+\Pd\Pubar$ and $\protect\sqrt{s}=200\GeV$
(taken from \protect\citere{racoonww_lep2res}a)}
\label{fi:ud_invmass}
\end{figure}

Figure~\ref{fi:ud_invmass} (l.h.s.) shows the invariant-mass distribution
of the $\Pd\Pubar$ quark pair for the semi-leptonic channel
$\Pep\Pem\to\nu_\mu\mu^+\Pd\Pubar$ at $\sqrt{s}=200\GeV$ at 
tree-level and with electroweak $\O(\alpha)$ corrections for two different
recombination cuts, $M_{\mathrm{rec}}=5$ and $25\GeV$.  
The recombination of photons with final-state charged fermions is
performed as described in\citere{racoonww_lep2res}: we first determine
the lowest invariant mass $M_{\gamma f}$ built by an emitted photon
and a charged final-state fermion. If $M_{\gamma f}$ is smaller than 
$M_{\mathrm{rec}}$, the photon momentum is
added to the one of the corresponding fermion $f$.  The maxima of the
corrected line shapes differ by up to \mbox{30--40}$\MeV$ for the two
values of $M_{\mathrm{rec}}$. As expected, there is a tendency to
shift the maxima to larger invariant masses if more and more photons
are recombined.  In Fig.~\ref{fi:ud_invmass} (r.h.s.) we display the
relative corrections 
$\delta=\rd\sigma/\rd\si_0-1$ for the two values of
$M_{\mathrm{rec}}$, which illustrates the strong dependence of the
corrected invariant-mass distributions on the treatment of the real
photons.  We obtain consistent results for the phase-space 
``slicing'' and the ``subtraction'' methods. The size of
the shown effects demonstrates that a careful treatment of real
photons is mandatory in the W-mass reconstruction at LEP2 accuracy.
\looseness -1

\begin{figure}
\setlength{\unitlength}{1cm}
{\centerline{
\begin{picture}(9,8.3)
\put(-.5,-1){\includegraphics{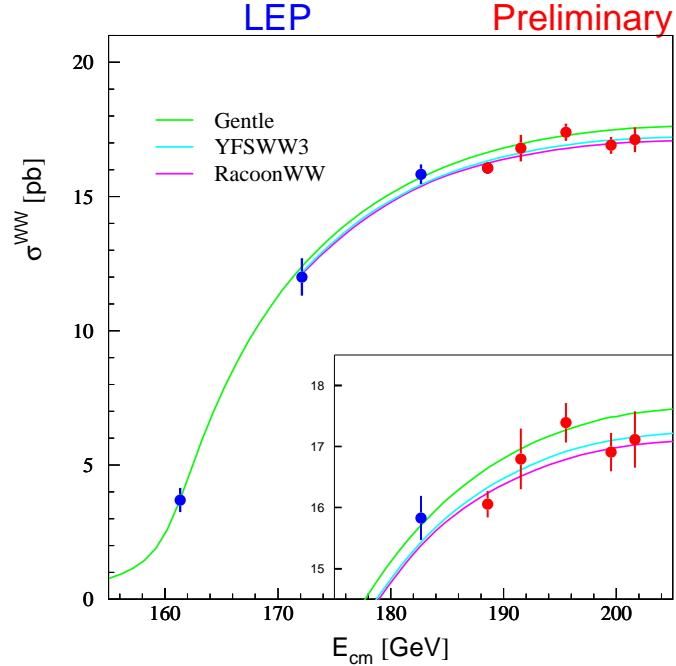}}
\end{picture}
} } 
\caption{Total WW production cross section at LEP2, as given by the
LEPEWWG \protect\cite{LEPEWWG}}
\label{fig:wwcs2000}
\end{figure}

Figure~\ref{fig:wwcs2000} shows a comparison of {\sc RacoonWW} results and 
of other predictions with recent LEP2 data, as given by the LEP Electroweak
Working Group \cite{lep2cs,LEPEWWG}.
The data are in good agreement with the predictions of {\sc RacoonWW}
and {\sc YFSWW3} \cite{yfsww}.
The predictions of these two generators differ between 0.5--0.7\%.%
\footnote{Meanwhile the dominant source of this difference has been
  found, and the new results of YFSWW3 are closer to the results of
  {\sc RacoonWW}. Details on the new YFSWW3 predictions can be found
  in \citere{lep2mcws}.}
More details on the conceptual differences of the two generators, as
well as a detailed comparison of numerical results, can be found in
\citere{lep2mcws}.
Figure~\ref{fig:wwcs2000} also includes the prediction provided by
{\sc GENTLE} \cite{gentle}, which differs from the {\sc RacoonWW} and
{\sc YFSWW3} results by 2--2.5\%. This difference is due to the neglect
of non-leading, non-universal ${\cal O}(\alpha)$ corrections in
{\sc GENTLE}. In summary, the comparison between SM predictions 
with the precise measurements of the W-pair production cross section
at LEP2 reveals evidence of non-leading electroweak radiative
corrections beyond the level of universal effects.

\end{document}